\documentclass{article}

     \usepackage[final]{neurips_2024}

\usepackage[utf8]{inputenc} 
\usepackage[T1]{fontenc}    
\usepackage{hyperref}       
\usepackage{url}            
\usepackage{booktabs}       
\usepackage{amsfonts}       
\usepackage{nicefrac}       
\usepackage{microtype}      
\usepackage{xcolor}         

 \usepackage{array,multirow,graphicx,amsmath}

\title{SELF-BART : A Transformer-based Molecular Representation Model using SELFIES}

\author{Indra Priyadarsini\\
  IBM Research - Tokyo\\ 
 \url{indra.ipd@ibm.com}  
  \And  
  Seiji Takeda\\
  IBM Research - Tokyo\\ 
  \url{seijitkd@jp.ibm.com}
  \And 
  Lisa Hamada \\
  IBM Research - Tokyo\\ 
  \url{lisa.hamada@ibm.com}
  \And 
  Emilio Vital Brazil \\
  IBM Research - Brazil\\ 
  \url{evital@br.ibm.com}
  \And 
  Eduardo Soares \\
  IBM Research - Brazil\\ 
  \url{eduardo.soares@ibm.com}
  \And 
  Hajime Shinohara \\
  IBM Research - Tokyo\\
  \url{hajime.shinohara1@ibm.com}
}

\begin{document}

\maketitle

\begin{abstract}
Large-scale molecular representation methods have revolutionized applications in material science, such as drug discovery, chemical modeling, and material design. With the rise of transformers, models now learn representations directly from molecular structures.  In this study, we develop an encoder-decoder model based on BART that is capable of leaning molecular representations and generate new molecules. Trained on SELFIES, a robust molecular string representation,  our model outperforms existing baselines in downstream tasks, demonstrating its potential in efficient and effective molecular data analysis and manipulation. 
\end{abstract}
\vspace{-3mm}
\section{Introduction}
Large-scale molecular representation methods are shown to be useful in various material science applications, such as virtual screening, drug discovery, chemical modeling, material design, and molecular dynamics simulations. With the progress in deep learning, numerous models have been developed to derive representations directly from molecular structures. Recently, transformer-based molecular representations have gained prominence in material informatics, offering significant potential for advancements in drug discovery, materials science, and related fields.  Recent works (\cite{chithrananda2020chemberta, bagal2021molgpt, ross2022large, chilingaryan2022bartsmiles, yuksel2023selformer}) have demonstrated the capability of transformer models in capturing complex relationships and patterns within molecular data with the help of attention mechanisms.
Most of these works are based on SMILES (Simplified Molecular Input Line Entry System) (\cite{weininger1988smiles}). However, one of the drawbacks of SMILES is that it does not guarantee syntactic and semantic validity of the molecule (\cite{krenn2020self}), thus leading to a possibility of learning invalid representations. SELFIES (SELF-referencing Embedded Strings) is another molecular string representation that was introduced by (\cite{krenn2020self}) to overcome the drawbacks of SMILES. Furthermore, in addition to achieving high accuracy predictions of molecular properties, a key objective within computational material informatics is to devise novel and functional molecules.
But most existing transformer models for material informatics are encoder-only models, which are not capable of generating new molecules.

In this paper, we introduce SELF-BART, a transformer-based model capable of capturing intricate molecular relationships and interactions. Unlike most existing works that utilize encoder-only models, we propose an encoder-decoder model based on BART (Bidirectional and Auto-Regressive Transformers) (\cite{lewis2019bart}). This model not only efficiently learns molecular representations but is also capable of auto-regressively generating new molecules  from these representations. This capability is particularly impactful for novel molecule design and generation, facilitating efficient and effective analysis and manipulation of molecular data.


\begin{figure*}
\begin{center}
  \includegraphics[width=0.7\textwidth]{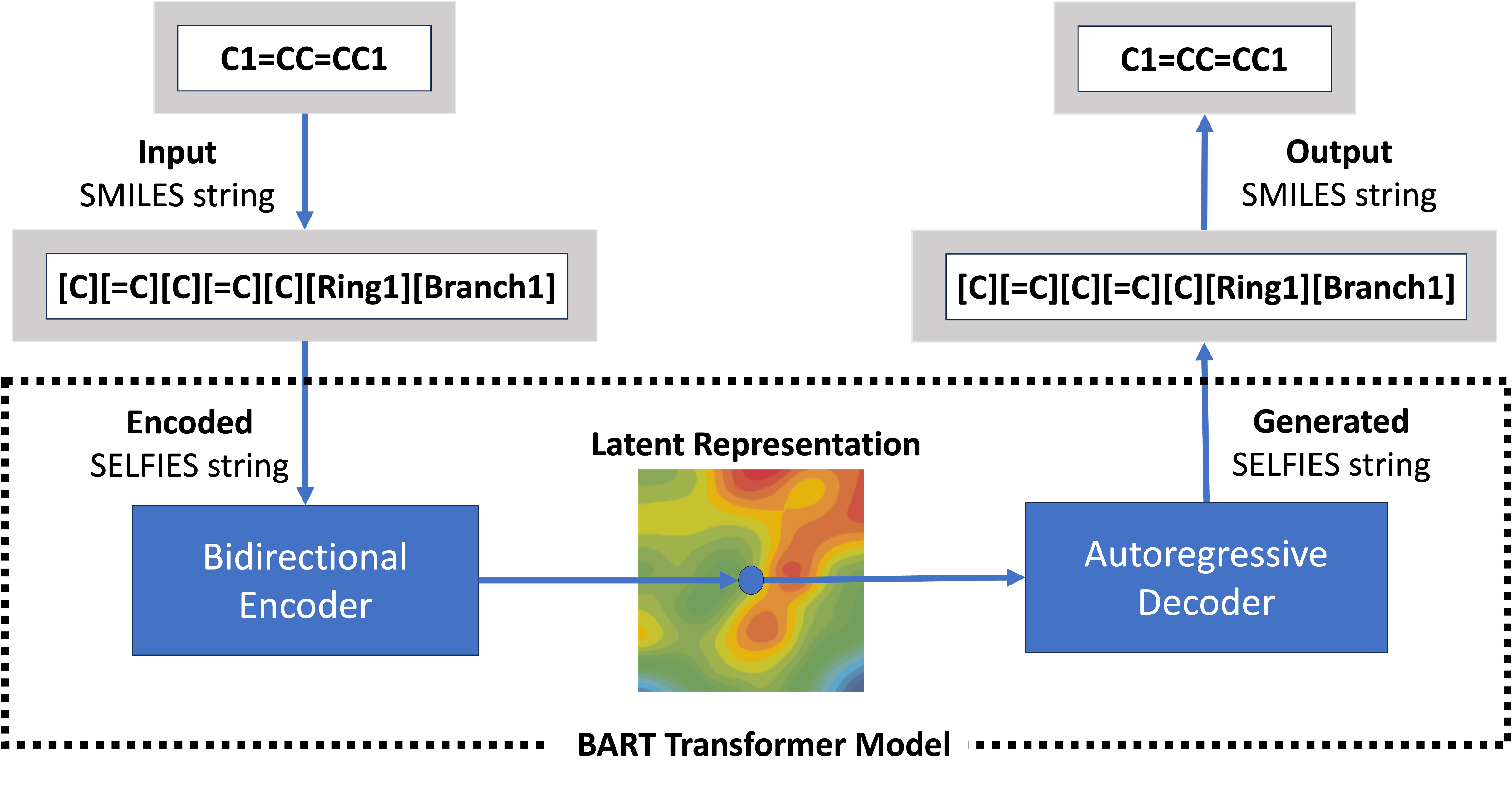}
\end{center}
\vspace{-2mm}
  \caption{Model architecture}
  \vspace{-2mm}
  \label{fig:model}
\end{figure*}

\vspace{-2mm}
\section{Model}
The proposed SELF-BART model is an encoder-decoder architecture derived from the BART (Bidirectional Auto-Regressive Transformer) model (\cite{lewis2019bart}). The encoder processes the sequence of input token bidirectionally and the decoder generates the sequence autoregressively. The SELF-BART model is trained using SELFIES as it  provides a more concise and interpretable representation, making it suitable for machine learning applications where compactness and generalization are important (\cite{krenn2020self}). During pre-training the model is trained with a denoising objective function. The model is trained using the ZINC-22 (\cite{tingle2023zinc}) and PubChem (\cite{kim2016pubchem}) datasets. The dataset consists of molecules represented in SMILES notation. We convert these SMILES strings to SELFIES  using the \textsc{selfies} API (\cite{krenn2020self}). In SELFIES each atom or bond is represented by symbols enclosed in \textbf{\lbrack ~\rbrack}, which are then tokenized using a word level tokenization scheme where each symbol or bond in \textbf{\lbrack ~\rbrack} is treated as a word. Further 15\% of the tokens are randomly masked and the model is trained using a denoising objective where the model learns to predict the next token in the original sequence, conditioned on both the corrupted sequence and the already decoded part of the original sequence.
The objective function is given as,
\[
\mathcal{L}_{\text{denoise}} = - \sum_{t=1}^{T} \log P(Y_t | Y_{<t}, X_{\text{corrupt}}; \theta)
\]
where, \( Y_t \) is the \( t \)-th token in the original sequence \( Y \), \( Y_{<t} \) represents the tokens preceding \( t \) in the target sequence, \( X_{\text{corrupt}} \) is the corrupted input sequence, \( \theta \) are the model parameters, and \( P(Y_t | Y_{<t}, X_{\text{corrupt}}; \theta) \) is the probability predicted by the model for token \( Y_t \), conditioned on the corrupted input and the previously generated tokens.
Figure 1 illustrates the pre-training model architecture. We hypothesize that the encoder-decoder structure of the SELF-BART model, combined with the denoising objective, provides better molecular representations. Moreover, training on SELFIES instead of SMILES ensures that the encoder output represents only valid molecules, enhancing the robustness of the molecular representations which are used for downstream tasks such as property prediction.

\begin{table*}[!htb]
\centering
\small
\begin{tabular}{llll}
\hline
\textbf{Dataset} &   \textbf{Description} & \textbf{\#Samples} & \textbf{Metric}  \\ \hline
BACE   & Binary labels on  $\beta$-secretase 1 (BACE1) binding properties & 1,513  & ROC-AUC \\ 
ClinTox   &  Binary labels on clinical toxicity data on FDA-approved drugs & 1478 & ROC-AUC\\ 
BBBP  & Binary labels on blood–brain barrier permeability & 2,039 & ROC-AUC \\ 
HIV   & Binary labels on the ability to inhibit HIV replication & 41,127  & ROC-AUC\\ 
SIDER   & Drug side effect classification for 27 types of adverse effects & 1,427 & ROC-AUC\\ 
Tox21   & Qualitative toxicity measurements on 12 targets & 7,831 & ROC-AUC
\\ 
Esol   & Water solubility prediction of small molecules & 1,128 & RMSE
\\Lipophilicity   & Prediction of octanol-water partition coefficient (logD) & 4,200 & RMSE
\\
Freesolv   & Hydration free energy of small molecules in water & 642 & RMSE
\\
\hline
\end{tabular}
\caption{Description of the benchmark datasets used in the evaluation of the proposed model.}
\vspace{-4mm}
\label{table:datasets}
\end{table*}
\vspace{-1mm}
\section{Results and Discussions}
\vspace{-1mm}
To evaluate the effectiveness of our proposed model on both molecular property prediction tasks and molecule generation tasks. For the molecule property predition tasks, we conducted evaluations using a comprehensive set of 9 distinct benchmark datasets sourced from MoleculeNet (\cite{wu2018moleculenet}). The details of the benchmarks used are illustrated in Table \ref{table:datasets}. We evaluate 6 datasets for the classification task and 3 datasets for regression tasks. To ensure a robust and unbiased assessment, we maintained consistency with the MoleculeNet benchmark by adopting identical train/validation/test splits for all tasks (\cite{wu2018moleculenet}). We compare the performance of the proposed SELF-BART model with various graph-based and text-based models. The SELF-BART model used in the evaluations is a 354M parameter model trained on 1B samples drawn from a combination of ZINC and PubChem datasets with a vocabulary of 3160 tokens. Futhermore, for the molecule generation tasks we conduct a preliminary analysis of the SELF-BART model and compare its results with existing molecular generative models.


\begin{table*}[b]
\centering
\small
\begin{tabular}{lcccccc}
\hline
\textbf{Model} & \textbf{BBBP} & \textbf{ClinTox} & \textbf{HIV} & \textbf{BACE} & \textbf{SIDER} & \textbf{Tox21} \\
\hline
RF (\cite{ross2022large}) & 71.4 & 71.3 & 78.1 & 86.7 & 68.4 & 76.9 \\
SVM (\cite{ross2022large}) & 72.9 & 66.9 & 79.2 & 86.2 & 68.2 & 81.8 \\
MGCN (\cite{lu2019molecular}) & 85.0 & 63.4 & 73.8 & 73.4 & 55.2 & 70.7 \\
D-MPNN (\cite{yang2019analyzing}) & 71.2 & 90.5 & 75.0 & 85.3 & 63.2 & 68.9 \\
DimeNet (\cite{gasteiger2020directional}) & - & 76.0 & - & - & 61.5 & 78.0 \\
Hu, et al. (\cite{hu2019strategies}) & 70.8 & 78.9 & 80.2 & 85.9 & 65.2 & 78.7 \\
N-Gram (\cite{liu2019n}) & 91.2 & 85.5 & 83.0 & 87.6 & 63.2 & 76.9 \\
MolCLR (\cite{wang2022molecular}) & 73.6 & 93.2 & 80.6 & 89.0 & 68.0 & 79.8 \\
GraphMVP (\cite{liu2021pre}) & 72.4 & 77.5 & 77.0 & 81.2 & 63.9 & 74.4 \\
GeomGCL (\cite{liu2021pre}) & - & 91.9 & - & - & 64.8 & 85.0 \\
GEM (\cite{fang2022geometry}) & 72.4 & 90.1 & 80.6 & 85.6 & 67.2 & 78.1 \\
\hline
ChemBerta (\cite{chithrananda2020chemberta}) & 64.3 & 73.3 & 62.2 & 79.9 & - & - \\
ChemBerta2 (\cite{ahmad2022chemberta}) & 71.94 & 90.7 & - & 85.1 & - & - \\
Galatica 30B (\cite{taylor2022galactica}) & 59.6 &  82.2 & 75.9 & 72.7 & 61.3 & 68.5 \\
Galatica 120B (\cite{taylor2022galactica}) & 66.1 &  82.6 & 74.5 & 61.7 &   63.2 & 68.9 \\
Uni-Mol (\cite{zhou2023uni}) & 72.9 &  91.9 & 80.8 & 85.7 &   65.9 & 79.6 \\
SELFormer (\cite{yuksel2023selformer}) & 90.2 & - & 68.1 & 83.2 & \textbf{74.5} & 65.3 \\
MoLFormer-XL (\cite{ross2022large}) & 93.7 & 94.8 & 82.2 & 88.2 & 69.0 & \textbf{84.7} \\

\hline
SELF-BART   & \textbf{95.2} & \textbf{96.9} & \textbf{83.0} & \textbf{89.3} & 65.0 & 76.5 \\
\hline
\end{tabular}
\caption{Results of the evaluation on classification tasks of MoleculeNet benchmark datasets}
\vspace{-5mm}
\label{classification}
\end{table*}

\vspace{-2mm}
\subsection{Molecular Property Prediction Tasks}
 We evaluated the SELF-BART models on nine benchmark from MoleculeNet \cite{wu2018moleculenet}. These tasks include four binary classification tasks using BACE, ClinTox, BBBP and HIV datasets, two multi-label classification task using SIDER and Tox21 datasets, and three regression tasks using the esol, freesolv and lipophilicity datasets.  For the evaluation, we used molecular embeddings generated by the SELF-BART models as input features. We use XGBoost (\cite{Chen:2016:XST:2939672.2939785}) as the downstream task model and Optuna (\cite{optuna_2019}) for hyperparameter tuning. The results corresponding to the optimal hyperparameters are reported. The performance is measured using the ROC-AUC and RMSE metrics. 
Table \ref{classification} presents the performance of the SELF-BART models compared to other molecular graph-based, geomentry-based models and molecular string-based models. ChemBERTa, Galatica, Uni-Mol and MolFormer are trained on SMILES representations, while SELFormer and the proposed SELF-BART model are trained on SELFIES representations. As shown in Table \ref{classification}, the SELF-BART  model outperforms the other models in four out of six tasks. We also evaluate the performance of the models on 3 regression task, the results of which are presented in Table \ref{regression}. The SELF-BART model outperforms the other models in two out of three tasks. The improved performance of SELF-BART can be attributed to encoder-decoder architecture of model being trained on SELFIES,  which ensures that the learned representations correspond to valid molecules. This approach substantially improves the robustness and quality of the molecular representations. Although both SELFormer and the proposed SELF-BART model are trained on SELFIES, SELF-BART demonstrates superior performance. This enhancement is primarily due to SELF-BART's encoder-decoder architecture combined with a denoising objective, in contrast to SELFormer's encoder-only architecture. This design choice significantly improves the robustness and quality of the molecular representations. 

\begin{table}
\small
\centering
\begin{tabular}{lccc}
\hline
\textbf{Model} & \textbf{ESOL} & \textbf{FreeSolv} & \textbf{Lipophilicity} \\
\hline
D-MPNN(\cite{yang2019analyzing}) & 1.050 & 2.082 & 0.683 \\
Hu et al.(\cite{hu2019strategies}) & 1.220 & 2.830 & 0.740 \\
MGCN(\cite{lu2019molecular}) & 1.270 & 3.350 & 1.110 \\
GEM(\cite{fang2022geometry}) & 0.798 & 1.877 & 0.660 \\
SchNet(\cite{schutt2017schnet}) & 1.050 & 3.220 & 0.910 \\
KPGT(\cite{li2022kpgt}) & 0.803 & 2.121 & \textbf{0.600} \\
GraphMVP-C(\cite{liu2021pre}) & 1.029 & - & 0.681 \\
GCN(\cite{kipf2016semi}) & 1.430 & 2.870 & 0.850 \\
GIN(\cite{xu2018powerful}) & 1.450 & 2.760 & 0.850 \\
MolCLR(\cite{wang2022molecular}) & 1.110 & 2.200 & 0.650 \\
ChemBERTa-2(\cite{ahmad2022chemberta})& - & - & 0.986 \\
MolFormer(\cite{ross2022large}) & 0.755 & 2.022 & 0.840 \\
SELFformer(\cite{yuksel2023selformer}) & 0.682 & 2.797 & 0.735 \\
\hline
SELF-BART  & \textbf{0.454} & \textbf{1.397} & 0.771 \\
\hline
\\
\end{tabular}
\caption{Results of the evaluation on regression tasks of MoleculeNet benchmark datasets}
\label{regression}
\vspace{-6mm}
\end{table}

\subsection{Molecule Generation Task}
The SELF-BART model is an encoder-decoder architecture, making it not only capable of providing robust molecular representations but also adept at generating molecules. In this section, we analyze the SELF-BART model's performance in non-conditioned molecular generation. Given the infinitely large and unexplored chemical space, it is crucial for a molecular generative model to understand molecular grammar and rules, ensuring the generation of novel and valid molecules. As a preliminary analysis, we evaluate the SELF-BART model's ability to generate molecules. For this purpose, we use the decoder, initializing it with the begin of sentence \texttt{<bos>} token to generate 10,000 molecules. This evaluation helps us understand the model's proficiency in producing diverse and valid molecular structures. The metrics we use in this analysis are validity, uniqueness, novelty and internal diversity. The metric scores are presented in Table \ref{table:comparison}. The metrics for CharRNN, VAE, AAE, LatentGAN, JT-VAE and MolGPT are values reported from (\cite{bagal2021molgpt}) trained on MOSES dataset, while SELF-BART was trained on 1B samples from ZINC-22 and PubChem. 
From the results, we can observe that the SELF-BART model is equally performant in generating unique, valid,
and novel molecules with the high internal diversity, thus confirming its effectiveness in generating molecules of varying structures and quality compared to similar baseline methods.



\begin{table*}[!htb]
\small 
\centering
\begin{tabular}{lccccc}
\hline
\textbf{Models} & \textbf{Validity} & \textbf{unique@10K} & \textbf{Novelty} & \textbf{IntDiv$_1$} & \textbf{IntDiv$_2$} \\ \hline
CharRNN   & 0.975 & 0.999 & 0.842 & 0.856 & 0.85 \\ 
VAE       & 0.977 & 0.998 & 0.695 & 0.856 & 0.85 \\ 
AAE       & 0.937 & 0.997 & 0.793 & 0.856 & 0.85 \\ 
LatentGAN & 0.897 & 0.997 & 0.949 & 0.857 & 0.85 \\ 
JT-VAE    & 1.0 & 0.999 & 0.914 & 0.855 & 0.849 \\ 
MolGPT    & 0.994 & 1.0 & 0.797 & 0.857 & 0.851 \\ \hline
SELF-BART    & 0.998 & 0.999 & 1.0 & 0.918 & 0.908 \\ \hline
\end{tabular}
\caption{Comparison of different models based on various metrics used in evaluating molecular generative models. }
\label{table:comparison}
\vspace{-2mm}
\end{table*}

\section{Conclusion}
This paper presents SELF-BART, an encoder-decoder transformer model designed to effectively learn representations of the chemical space. By training on SELFIES strings, SELF-BART ensures the validity of the molecules during pre-training, which enhances the robustness of its molecular representations. The model's effectiveness is demonstrated through performance evaluations on benchmark classification and regression tasks from MoleculeNet. The SELF-BART model achieved state-of-the-art results in most tasks. Although the primary focus is on molecular representation for downstream tasks, we provided an initial exploration of the model's ability to generate molecules without conditioning. The preliminary analysis showed that the model was capable of generating valid and novel molecules with good structural diversity. Future work will investigate the model's generative capabilities further, including conditioned molecular generation, and examine its performance with scaling and conditioned generative modeling.

\bibliography{main}

\begin{thebibliography}{28}
\providecommand{\natexlab}[1]{#1}
\providecommand{\url}[1]{\texttt{#1}}
\expandafter\ifx\csname urlstyle\endcsname\relax
  \providecommand{\doi}[1]{doi: #1}\else
  \providecommand{\doi}{doi: \begingroup \urlstyle{rm}\Url}\fi

\bibitem[Ahmad et~al.(2022)Ahmad, Simon, Chithrananda, Grand, and
  Ramsundar]{ahmad2022chemberta}
Walid Ahmad, Elana Simon, Seyone Chithrananda, Gabriel Grand, and Bharath
  Ramsundar.
\newblock Chemberta-2: Towards chemical foundation models.
\newblock \emph{arXiv preprint arXiv:2209.01712}, 2022.

\bibitem[Akiba et~al.(2019)Akiba, Sano, Yanase, Ohta, and Koyama]{optuna_2019}
Takuya Akiba, Shotaro Sano, Toshihiko Yanase, Takeru Ohta, and Masanori Koyama.
\newblock Optuna: A next-generation hyperparameter optimization framework.
\newblock In \emph{Proceedings of the 25th {ACM} {SIGKDD} International
  Conference on Knowledge Discovery and Data Mining}, 2019.

\bibitem[Bagal et~al.(2021)Bagal, Aggarwal, Vinod, and
  Priyakumar]{bagal2021molgpt}
Viraj Bagal, Rishal Aggarwal, PK~Vinod, and U~Deva Priyakumar.
\newblock Molgpt: molecular generation using a transformer-decoder model.
\newblock \emph{Journal of Chemical Information and Modeling}, 62\penalty0
  (9):\penalty0 2064--2076, 2021.

\bibitem[Chen and Guestrin(2016)]{Chen:2016:XST:2939672.2939785}
Tianqi Chen and Carlos Guestrin.
\newblock {XGBoost}: A scalable tree boosting system.
\newblock In \emph{Proceedings of the 22nd ACM SIGKDD International Conference
  on Knowledge Discovery and Data Mining}, KDD '16, pages 785--794, New York,
  NY, USA, 2016. ACM.
\newblock ISBN 978-1-4503-4232-2.
\newblock \doi{10.1145/2939672.2939785}.
\newblock URL \url{http://doi.acm.org/10.1145/2939672.2939785}.

\bibitem[Chilingaryan et~al.(2022)Chilingaryan, Tamoyan, Tevosyan, Babayan,
  Khondkaryan, Hambardzumyan, Navoyan, Khachatrian, and
  Aghajanyan]{chilingaryan2022bartsmiles}
Gayane Chilingaryan, Hovhannes Tamoyan, Ani Tevosyan, Nelly Babayan, Lusine
  Khondkaryan, Karen Hambardzumyan, Zaven Navoyan, Hrant Khachatrian, and Armen
  Aghajanyan.
\newblock Bartsmiles: Generative masked language models for molecular
  representations.
\newblock \emph{arXiv preprint arXiv:2211.16349}, 2022.

\bibitem[Chithrananda et~al.(2020)Chithrananda, Grand, and
  Ramsundar]{chithrananda2020chemberta}
Seyone Chithrananda, Gabriel Grand, and Bharath Ramsundar.
\newblock Chemberta: large-scale self-supervised pretraining for molecular
  property prediction.
\newblock \emph{arXiv preprint arXiv:2010.09885}, 2020.

\bibitem[Fang et~al.(2022)Fang, Liu, Lei, He, Zhang, Zhou, Wang, Wu, and
  Wang]{fang2022geometry}
Xiaomin Fang, Lihang Liu, Jieqiong Lei, Donglong He, Shanzhuo Zhang, Jingbo
  Zhou, Fan Wang, Hua Wu, and Haifeng Wang.
\newblock Geometry-enhanced molecular representation learning for property
  prediction.
\newblock \emph{Nature Machine Intelligence}, 4\penalty0 (2):\penalty0
  127--134, 2022.

\bibitem[Gasteiger et~al.(2020)Gasteiger, Gro{\ss}, and
  G{\"u}nnemann]{gasteiger2020directional}
Johannes Gasteiger, Janek Gro{\ss}, and Stephan G{\"u}nnemann.
\newblock Directional message passing for molecular graphs.
\newblock \emph{arXiv preprint arXiv:2003.03123}, 2020.

\bibitem[Hu et~al.(2019)Hu, Liu, Gomes, Zitnik, Liang, Pande, and
  Leskovec]{hu2019strategies}
Weihua Hu, Bowen Liu, Joseph Gomes, Marinka Zitnik, Percy Liang, Vijay Pande,
  and Jure Leskovec.
\newblock Strategies for pre-training graph neural networks.
\newblock \emph{arXiv preprint arXiv:1905.12265}, 2019.

\bibitem[Kim et~al.(2016)Kim, Chen, Gindulyte, He, He, Shoemaker, Thiessen,
  Bolton, Fu, Han, et~al.]{kim2016pubchem}
Sunghwan Kim, Jie Chen, Asta Gindulyte, Jane He, Siqian He, Benjamin~A
  Shoemaker, Paul~A Thiessen, Evan~E Bolton, Gang Fu, Lianyi Han, et~al.
\newblock Pubchem substance and compound databases.
\newblock \emph{Nucleic acids research}, 44\penalty0 (D1):\penalty0
  D1202--D1213, 2016.

\bibitem[Kipf and Welling(2016)]{kipf2016semi}
Thomas~N Kipf and Max Welling.
\newblock Semi-supervised classification with graph convolutional networks.
\newblock \emph{arXiv preprint arXiv:1609.02907}, 2016.

\bibitem[Krenn et~al.(2020)Krenn, H{\"a}se, Nigam, Friederich, and
  Aspuru-Guzik]{krenn2020self}
Mario Krenn, Florian H{\"a}se, AkshatKumar Nigam, Pascal Friederich, and Alan
  Aspuru-Guzik.
\newblock Self-referencing embedded strings (selfies): A 100\% robust molecular
  string representation.
\newblock \emph{Machine Learning: Science and Technology}, 1\penalty0
  (4):\penalty0 045024, 2020.

\bibitem[Lewis et~al.(2019)Lewis, Liu, Goyal, Ghazvininejad, Mohamed, Levy,
  Stoyanov, and Zettlemoyer]{lewis2019bart}
Mike Lewis, Yinhan Liu, Naman Goyal, Marjan Ghazvininejad, Abdelrahman Mohamed,
  Omer Levy, Ves Stoyanov, and Luke Zettlemoyer.
\newblock Bart: Denoising sequence-to-sequence pre-training for natural
  language generation, translation, and comprehension.
\newblock \emph{arXiv preprint arXiv:1910.13461}, 2019.

\bibitem[Li et~al.(2022)Li, Zhao, and Zeng]{li2022kpgt}
Han Li, Dan Zhao, and Jianyang Zeng.
\newblock Kpgt: knowledge-guided pre-training of graph transformer for
  molecular property prediction.
\newblock In \emph{Proceedings of the 28th ACM SIGKDD Conference on Knowledge
  Discovery and Data Mining}, pages 857--867, 2022.

\bibitem[Liu et~al.(2019)Liu, Demirel, and Liang]{liu2019n}
Shengchao Liu, Mehmet~F Demirel, and Yingyu Liang.
\newblock N-gram graph: Simple unsupervised representation for graphs, with
  applications to molecules.
\newblock \emph{Advances in neural information processing systems}, 32, 2019.

\bibitem[Liu et~al.(2021)Liu, Wang, Liu, Lasenby, Guo, and Tang]{liu2021pre}
Shengchao Liu, Hanchen Wang, Weiyang Liu, Joan Lasenby, Hongyu Guo, and Jian
  Tang.
\newblock Pre-training molecular graph representation with 3d geometry.
\newblock \emph{arXiv preprint arXiv:2110.07728}, 2021.

\bibitem[Lu et~al.(2019)Lu, Liu, Wang, Huang, Lin, and He]{lu2019molecular}
Chengqiang Lu, Qi~Liu, Chao Wang, Zhenya Huang, Peize Lin, and Lixin He.
\newblock Molecular property prediction: A multilevel quantum interactions
  modeling perspective.
\newblock In \emph{Proceedings of the AAAI conference on artificial
  intelligence}, volume~33, pages 1052--1060, 2019.

\bibitem[Ross et~al.(2022)Ross, Belgodere, Chenthamarakshan, Padhi, Mroueh, and
  Das]{ross2022large}
Jerret Ross, Brian Belgodere, Vijil Chenthamarakshan, Inkit Padhi, Youssef
  Mroueh, and Payel Das.
\newblock Large-scale chemical language representations capture molecular
  structure and properties.
\newblock \emph{Nature Machine Intelligence}, 4\penalty0 (12):\penalty0
  1256--1264, 2022.

\bibitem[Sch{\"u}tt et~al.(2017)Sch{\"u}tt, Kindermans, Sauceda~Felix, Chmiela,
  Tkatchenko, and M{\"u}ller]{schutt2017schnet}
Kristof Sch{\"u}tt, Pieter-Jan Kindermans, Huziel~Enoc Sauceda~Felix, Stefan
  Chmiela, Alexandre Tkatchenko, and Klaus-Robert M{\"u}ller.
\newblock Schnet: A continuous-filter convolutional neural network for modeling
  quantum interactions.
\newblock \emph{Advances in neural information processing systems}, 30, 2017.

\bibitem[Taylor et~al.(2022)Taylor, Kardas, Cucurull, Scialom, Hartshorn,
  Saravia, Poulton, Kerkez, and Stojnic]{taylor2022galactica}
Ross Taylor, Marcin Kardas, Guillem Cucurull, Thomas Scialom, Anthony
  Hartshorn, Elvis Saravia, Andrew Poulton, Viktor Kerkez, and Robert Stojnic.
\newblock Galactica: A large language model for science.
\newblock \emph{arXiv preprint arXiv:2211.09085}, 2022.

\bibitem[Tingle et~al.(2023)Tingle, Tang, Castanon, Gutierrez, Khurelbaatar,
  Dandarchuluun, Moroz, and Irwin]{tingle2023zinc}
Benjamin~I Tingle, Khanh~G Tang, Mar Castanon, John~J Gutierrez, Munkhzul
  Khurelbaatar, Chinzorig Dandarchuluun, Yurii~S Moroz, and John~J Irwin.
\newblock Zinc-22 - a free multi-billion-scale database of tangible compounds
  for ligand discovery.
\newblock \emph{Journal of chemical information and modeling}, 63\penalty0
  (4):\penalty0 1166--1176, 2023.

\bibitem[Wang et~al.(2022)Wang, Wang, Cao, and
  Barati~Farimani]{wang2022molecular}
Yuyang Wang, Jianren Wang, Zhonglin Cao, and Amir Barati~Farimani.
\newblock Molecular contrastive learning of representations via graph neural
  networks.
\newblock \emph{Nature Machine Intelligence}, 4\penalty0 (3):\penalty0
  279--287, 2022.

\bibitem[Weininger(1988)]{weininger1988smiles}
David Weininger.
\newblock Smiles, a chemical language and information system. 1. introduction
  to methodology and encoding rules.
\newblock \emph{Journal of chemical information and computer sciences},
  28\penalty0 (1):\penalty0 31--36, 1988.

\bibitem[Wu et~al.(2018)Wu, Ramsundar, Feinberg, Gomes, Geniesse, Pappu,
  Leswing, and Pande]{wu2018moleculenet}
Zhenqin Wu, Bharath Ramsundar, Evan~N Feinberg, Joseph Gomes, Caleb Geniesse,
  Aneesh~S Pappu, Karl Leswing, and Vijay Pande.
\newblock Moleculenet: a benchmark for molecular machine learning.
\newblock \emph{Chemical science}, 9\penalty0 (2):\penalty0 513--530, 2018.

\bibitem[Xu et~al.(2018)Xu, Hu, Leskovec, and Jegelka]{xu2018powerful}
Keyulu Xu, Weihua Hu, Jure Leskovec, and Stefanie Jegelka.
\newblock How powerful are graph neural networks?
\newblock \emph{arXiv preprint arXiv:1810.00826}, 2018.

\bibitem[Yang et~al.(2019)Yang, Swanson, Jin, Coley, Eiden, Gao, Guzman-Perez,
  Hopper, Kelley, Mathea, et~al.]{yang2019analyzing}
Kevin Yang, Kyle Swanson, Wengong Jin, Connor Coley, Philipp Eiden, Hua Gao,
  Angel Guzman-Perez, Timothy Hopper, Brian Kelley, Miriam Mathea, et~al.
\newblock Analyzing learned molecular representations for property prediction.
\newblock \emph{Journal of chemical information and modeling}, 59\penalty0
  (8):\penalty0 3370--3388, 2019.

\bibitem[Y{\"u}ksel et~al.(2023)Y{\"u}ksel, Ulusoy, {\"U}nl{\"u}, and
  Do{\u{g}}an]{yuksel2023selformer}
Atakan Y{\"u}ksel, Erva Ulusoy, Atabey {\"U}nl{\"u}, and Tunca Do{\u{g}}an.
\newblock Selformer: molecular representation learning via selfies language
  models.
\newblock \emph{Machine Learning: Science and Technology}, 4\penalty0
  (2):\penalty0 025035, 2023.

\bibitem[Zhou et~al.(2023)Zhou, Gao, Ding, Zheng, Xu, Wei, Zhang, and
  Ke]{zhou2023uni}
Gengmo Zhou, Zhifeng Gao, Qiankun Ding, Hang Zheng, Hongteng Xu, Zhewei Wei,
  Linfeng Zhang, and Guolin Ke.
\newblock Uni-mol: A universal 3d molecular representation learning framework.
\newblock In \emph{The Eleventh International Conference on Learning
  Representations}, 2023.
\newblock URL \url{https://openreview.net/forum?id=6K2RM6wVqKu}.

\end{thebibliography}
\bibliographystyle{plainnat}

\end{document}